# THE POTENTIAL OF THE LINAC-RING TYPE COLLIDERS FOR PARTICLE AND NUCLEAR PHYSICS

A. K. Çiftçi, E. Recepoğlu, Dept. of Physics, Fac. of Science, Ankara Univ., Ankara, TURKEY
S. Sultansoy, Dept. of Physics, Fac. of Art and Sciences., Gazi Univ., Ankara, TURKEY
Ö. Yavaş, Dept. of Eng. Physics, Fac. of Engineering, Ankara Univ., Ankara, TURKEY
M. Yılmaz, Dept. of Physics, Fac. of Art and Sciences, Gazi University, Ankara, TURKEY

*Abstract*

Linac-ring type colliders will open new windows for both energy frontier and particle factories. Concerning the first direction, these machines seem to be a sole way to TeV scale in lepton-hadron collisison at constituent level. An essential advantage of the linac-ring type lepton-hadron colliders is the possibility of the construction of $\gamma p$, $\gamma A$ and FEL$\gamma$-A colliders based on them. Today, eRHIC, THERA (TESLA on HERA) and Linac*LHC can be considered as realistic candidates for future lepton-hadron and photon hadron colliders. When it comes to factories, one can reach essentially higher luminosities comparing to standard ring-ring type machines. For example, $L=10^{34}$ cm$^{-2}$s$^{-1}$ can be achieved for phi and charm-tau factories. In this presentation we briefly discuss the parameters and physics search potential of the linac-ring type machines.

## 1 INTRODUCTION

There are two well-known ways to reach TeV scale c.m. energies at constituent level: the ring-ring type proton-proton colliders and the linear $e^+e^-$ colliders (including $\gamma e$ and $\gamma\gamma$ options [1]). Recently two more methods have been added to the list, namely: the ring type $\mu^+\mu^-$ colliders [2] and the linac-ring type ep and eA colliders [3] (including $\gamma p$, $\gamma A$ and FEL$\gamma$A options [4-6]).

The investigation of physics phenomena at extreme small x but sufficiently high $Q^2$ is very important for understanding the nature of strong interactions at all levels from nucleus to partons. At the same time, the results from lepton-hadron collidres are necessary for adequate interpretation of physics at future hadron colliders. Today, linac-ring type ep machines and additional options based on them seem to be the main way to TeV scale in lepton-hadron and photon-hadron collisions.

On the other hand, the idea of colliding of the electron beam from a linac with the positron beam stored in a ring is widely discussed during the last decade in the context of high luminosity particle facrories, namely, B factory, $\phi$ factory, c-$\tau$ factory etc.

## 2 THERA BASED COLLIDERS

The opportunity to collide electron and photon beams from TESLA with proton and nucleus beams in HERA were taken into into account by choosing the direction of TESLA tangential to HERA [7].

Main parameters of three possible options for THERA based ep collider are presented in Table 1.

Table 1. Main parameters of THERA based ep collider.

|  | Option 1 | Option 2 | Option 3 |
|---|---|---|---|
| $E_e$, TeV | 0.25 | 0.5 | 0.8 |
| $E_p$, TeV | 1.0 | 0.5 | 0.8 |
| $\sqrt{s}$, TeV | 1.0 | 1.0 | 1.6 |
| L, $10^{31}$ cm$^{-2}$s$^{-1}$ | 0.4 | 2.5 | 1.6 |

A brief account of some SM physics topics (structure functions, hadronic final states, high $Q^2$ region etc.) is presented below [7]:
- The extension of kinematic range down to x~$10^{-6}$ allows access to the high parton-density domain and its detailed exploration in the deep-inelastic regime.
- The measurement of proton structure functions at THERA will be essential for determining quark and gluon distributions in the proton in an unexplored kinematic region.
- The study of forward-going jets at THERA is expected to reveal the mechanism for the evolution of QCD radiation at low x.
- The total cross sections for charm and beauty production are expected to increase by factors of three and five, respectively, as compared to HERA.
- THERA will operate beyond the electroweak unification scale and is thus a trully "electroweak interaction machine".
- THERA will probe physics beyond the Standard Model.

Refering for details to [4] and [8] let us note that luminosity of $\gamma p$ collisions is approximately one half of the luminosity of ep collisions (photon beam is obtained due to Compton backscattering of laser beam on electron beam from TESLA). A partial list of physics goals of THERA based $\gamma p$ collider is following [9,10]:

- Total cross section at TeV scale, which can be extrapolated from existing low energy data as $\sigma(\gamma p \to hadrons) \approx 100 - 200 \mu b$.
- Two-jets events, $10^4$ events per working year with $p_t > 100$.
- Heavy quark pairs, $10^7$-$10^8$ ($10^6$-$10^7$, $10^2$-$10^3$) events per working year for $c\bar{c}(b\bar{b}, t\bar{t})$ pair production.
- Hadronic structure of the photon.
- Single W production, $10^4$-$10^5$ events per working year.
- Excited quarks ($u^*$ and $d^*$) with $m \leq 0.5$ TeV.
- Single leptoquarks with $m \leq 0.5$ TeV.
- Associate wino-squark and gluino-squark production if the sum of their masses is less than 0.5 TeV.

For the THERA based eA collider, the main limitation comes from fast emittance growth due to intra beam scattering, which is approximately proportional $(Z^2/A)^2(\gamma_A)^{-3}$. Nevertheless, sufficiently high luminosity can be achieved at least for light nuclei. For example, $L_{eC}=1.1 \cdot 10^{29}$cm$^{-2}$s$^{-1}$ for collisions of 300 GeV energy electron beams and Carbon beam with $n_C=8 \cdot 10^9$ and $\varepsilon_C^N=1.25\pi$·mrad·mm [11]. This value corresponds to $L_{int} \cdot A \approx 10 pb^{-1}$ per working year ($10^7$ s) needed from the physics point of view [12].

In our opinion, $\gamma A$ collider is the most promising option of the TESLA⊗HERA complex, because it will give unique opportunity to investigate small $x_g$ region in nuclear medium. Indeed, due to the advantage of the real $\gamma$ spectrum heavy quarks will be produced via $\gamma g$ fusion at characteristic

$$x_g \approx \frac{4m^2_{c(b)}}{0.83 \times (Z/A) \times s_{ep}},$$

which is approximately $(2 \div 3) \cdot 10^{-5}$ for charmed hadrons. Some of the physics goals of THERA based $\gamma A$ collider are listed below [9,10]:
- Total cross-section to clarify real mechanism of very high-energy $\gamma$–nucleus interactions.
- Investigation of hadronic structure of the proton in nuclear medium.
- According to the VMD, proposed machine will also be a $\rho$-nucleus collider.
- Formation of quark-gluon plasma at very high temperatures but relatively low nuclear density.
- Gluon distribution at extremely small $x_g$ in nuclear medium ($\gamma A \to QQ + X$)
- Investigation of both heavy quark and nuclear medium properties ($\gamma A \to J/\Psi(Y) + X$, $J/\Psi(Y) \to l^+l^-$).
- Existence of multi-quark clusters in nuclear medium and a few-nucleon correlation.

## 3 LinacxLHC BASED COLLIDERS

The center of mass energies which will be achieved at differenet options of this machine are an order larger than those at HERA and ~3 times larger than the energy region of THERA and LEP*LHC. In principle, luminosity values are ~7 times higher than those of corresponding options of HERA complex due to higher energy of protons.

Center of mass energy and luminosity for ep option are [13] $\sqrt{s} = 5.29$ TeV and $L_{ep} = 10^{32}$ cm$^{-2}$s$^{-1}$, recpectively; and additional factor of 3-4 can be provided by the "dynamic" focusing scheme [14]. A further increase will require cooling at injector stages. This machine, which will extend both the Q$^2$-range and x-range by more than two order of magnitude comparing to those explored by HERA, has a strong potential for both SM and BSM research. Few examples are: the discovery limit for the first generation leptoquarks is $m \approx 3$ TeV; the discovery limit for SUSY particles is $m_{\tilde{l}} + m_{\tilde{q}} \approx 1.5$ TeV and covers all six SUGRA points, which are used for SUSY analyses at LHC [15]; excited electron will be copiously produced up to $m_{e^*} \approx 3.5$ TeV etc.

In the case of $\gamma p$ option, the advantage in spectrum of backscattered photons and sufficiently high luminosity (for details see [4,13]), $L_{\gamma p} = 10^{32}$ cm$^{-2}$s$^{-1}$ at z = 0, will clearly manifest itself in searching of different phenomena.
The physics search potential of $\gamma p$ colliders is reviewed in [9]. The $\gamma p$ option will essentially enlarge the capacity of the Linac*LHC complex. For example, thousands of di-jets with $p_t >500$ GeV and hundred thousands of single W bosons will be produced, hundred millions of $b\bar{b}$ and $c\bar{c}$ pairs will give opportunity to explore the region of extremely small $x_g$ (~$10^{-6}$-$10^{-7}$) etc. Concerning the BSM physics:
- linac-ring type $\gamma p$ colliders are ideal are machines for u*, d*, and $Z_8$ search (the discovery limits are $m_{u^*} = 5$ TeV, $m_{d^*} = 4$ TeV, $m_{Z_8} = 4$ TeV)
- the fourth SM family quarks (the discovery limits are $m_{u_4} \approx 1$ TeV and $m_{d_4} \approx 0.8$ TeV) will be copiously produced, since their masses are predicted to be in the region 300÷700 GeV with preferable value $m_{u_4} \approx m_{d_4} \approx 640$ GeV (see [16] and references therein)
- because of parameter inflation, namely more than 160 observable free parameters in the three family MSSM (see [17] and references therein), SUSY should be realized at preonic level. Nevertheless, the discovery limits for different channels are: $m_{\tilde{w}} + m_{\tilde{q}} \approx 1.8$ TeV, $m_{\tilde{g}} + m_{\tilde{q}} \approx 1.6$ TeV, $m_{\tilde{\gamma}} + m_{\tilde{q}} \approx 0.5$ TeV, $m_{\tilde{q}} + m_{\tilde{q}} \approx 1.5$ TeV.

In the case of LHC nucleus beam, IBS effects in main ring are not crucial because of large value of $\gamma_A$. The main principal limitation for heavy nuclei coming from beam-beam tune shift may be weakened using flat beams at collision point. Rough estimations show that $L_{eA} \cdot A > 10^{31} cm^{-2} s^{-1}$ can be achieved at least for light and medium nuclei [11]. For electron-carbon and electron-Pb collisions, one has $L_{eC} \cdot A = 10^{31} cm^{-2} s^{-1}$ and $L_{ePb} \cdot A = 1.2 \cdot 10^{30} cm^{-2} s^{-1}$, correspondingly. This machine will extend both the $Q^2$-range and x-range by more than two orders of magnitude with respect to the region, which can be explored by HERA based eA collider.

In the case of γA collider, limitation on luminosity due to beam-beam tune shift is removed in the scheme with deflection of electron beam after conversion [4]. As it is shown in [13], $L_{\gamma C} \cdot A = 0.8 \cdot 10^{31} cm^{-2} s^{-1}$ and $L_{\gamma Pb} \cdot A = 10^{30} cm^{-2} s^{-1}$ at z = 5m. Center of mass energy of Linac*LHC based γA collider corresponds to $E_\gamma \sim$ PeV in the lab system. At this energy range, cosmic ray experiments have a few events per year, whereas γ-nucleus collider will give few billions events. This machine has a wide research capacity. Especially, the investigation of gluon distribution at extremely small $x_g$ will give a crucial information for QCD in nuclear medium.

## 5. LINAC-RING TYPE PARTICLE FACTORIES

### 5.1. Phi Factory

It is well known that, center of mass energy and luminosity are two important parameters from the point of view of particle physics. In order to make a linac-ring type $\phi$ factory feasible, its luminosity should exceed $10^{33}$ cm$^{-2}$s$^{-1}$. We have presented sets of parameters for a linac-ring type $\phi$ factory in [18]. About $4 \cdot 10^{11}$ events per working year ($10^7$ s) is expected. For main parameters and rates of $\phi$ decays at a linac-ring type $\phi$ factory see ref. [19].

### 5.2. Charm Factory

Recently CLEO-c [20] has been approved in order to explore the charm sector starting early 2003. Expected machine performance will be $L=3 \cdot 10^{32} cm^{-2} s^{-1}$ at $\sqrt{s} = 3.77$ GeV. We present proposed parameters for linac-ring type charm factory [21] and show that $L=10^{33 \div 34} cm^{-2} s^{-1}$ can be achieved. Expected numbers of $\Psi(3S)$ per working year is about $10^{10}$. Let us mentioned that $D\bar{D}$ mode is the dominant one for $\Psi(3S)$ decays. An additional advantage of the proposed charm factory is the asymmetrc kinematics. This feature will be important in investigations of $D^0\bar{D}^0$ oscillations and CP-violation in charmed particle decays.

### 5.3. Tau Factory

The maximum cross section of the process $e^+e^- \to \tau^+\tau^-$ is $\sigma = 3.56$ nb at $\sqrt{s} \approx 4.2$ GeV. In diference from $\phi$ and charm factories, in the case of $\tau$ factory we have consider the symmetric option ($E_{e^-} = E_{e^+} = 2.1$ GeV). Proposed set of parameters can be found in [21]. One can see that linac-ring type $\tau$ factory will produce $\sim 4 \cdot 10^8 \tau^+\tau^-$ pair working year, which exceeds by two order the statistics obtained at LEP and CLEO up to now.

## ACKNOWLEDGEMENTS

This work is supported by Turkish State Planning Organization under the Grant No DPT-2002K-120250.